# Terrestrial planetary dynamics: a view from U, Th geochemistry

Xuezhao Bao[*]

Department of Earth Sciences, University of Western Ontario

34-534, Platt's Lane, London, Ontario, Canada N6G 3A8

**Abstract**

The migration of U and Th inside a planet is controlled by its oxidation state imposed by the volatile composition. In the deep interior of a planet, an absence of oxidative volatiles will cause U and Th to stay in a state of metal or low valance compounds with a big density. Consequently, they migrate to the bottom of its mantle first, and then are gradually sequestered to its liquid metal core. Earth is rich in oxidative volatiles including water, therefore, U and Th in the core can be moved up by an internal circulation system consisting of the outer core, hot super plumes, asthenosphere and subduction zone (or cold super plumes). This internal circulation system is the key for the formation of plate tectonics, the geodynamo and the consequent geomagnetic field. Moreover, plentiful oxidative volatiles and water within Earth is the precondition to form such a circulation system. In the early stage (> 4 Ga), Mars developed an Earth-like internal circulation system due to relatively large amount of oxidative volatile compositions coming from its building material. This would have produced a dynamo and correspondingly an Earth-like magnetic field. However, this internal circulation system was destroyed by one or several giant impact events in the early stage, which drove off these volatile compositions. These events also shaped the striking hemispheric dichotomy structure on the Martian surface. The other result is that its dynamo and geomagnetic field have also disappeared. Since then, Mars has been the same as Mercury and Venus in that the heat release from the U and Th in their cores can not be moved by an internal circulation system gently, but by sporadically catastrophic resurfacing events (Venus), or super plumes (Mars) or gradual heat conduction (Mercury).

---

[*] Corresponding author. 1-519-673-0626; Email: xuezhaobao@hotmail.com

1. INTRODUCTION

The recent observations of geoneutrinos produced from the decay of U and Th indicate that the heat output from these radioactive elements within Earth's interior has an upper limit of 60 TW with a centre value of 16 TW (Araki, et al., 2005). Through a combination of measurements and calculations, Earth's surface heat flux output is ~44 TW (Pollack et al., 1993; Stein, 1995), and of this, the heat output of the continental crust is only ~8 TW (Stein, 1995, Korenaga, 2003). If the mantle is homogenous as required by whole mantle convection and as supported by seismic tomography (Ishida et al., 1999), using the concentration of U (8 ppb), Th (32 ppb) and K (100 ppm) from MORB (Mid-Oceanic Ridge Basalt) as the average concentrations for the mantle, then the heat output of the mantle is only 6 TW (e.g. Korenaga, 2003). The contribution to the total heat flux from secular heat is limited to the first Ga after Earth's formation based on the studies of Van den Berg and Yuen (2002), and Van den Berg et al. (2002). Therefore, the heat source of the remaining 30 TW must reside elsewhere within the Earth, but so far geophysical methods have not been able to detect it. The missing heat source(s) has (have) been called the heat source paradox (e.g. Korenaga, 2003).

The experimental data obtained in our previous studies (Bao et al., 2006; Bao and Secco, 2006) support that there is some amount of U in the core. However, using the highest extrapolated results of 10 ppb U in the core based on current primitive mantle U concentration value of 20 ppb (Bao and Secco, 2006), the heat energy from U in the core is less than 2 TW. Therefore, if radiogenic heat is in fact the main component of the missing heat source within Earth, then the real U, Th or K concentrations within the Earth's interior may be much higher than those assumed by the current geochemical model, which is based on Cl chondritic meteorites as Earth's building material (McDonough, 2003). However, as pointed out by Righter and Drake (2006), Earth's building material must be very different from the extant meteorites. Therefore, many of the calculations that are based on the current geochemical model, including the U concentration in the core as discussed above, may not be supported. On the other hand, the migration behavior of U and Th and their distribution in a planetary interior can be understood by high pressure (P) and temperature (T) experiments and theoretical analysis. Therefore, based on previous studies (Bao and Zhang, 1998; Bao, 1999), and the new high pressure and temperature experimental results in our previous study (Bao et al., 2006; Bao and Secco, 2006), the U and Th geochemistry is discussed in this paper along with their implications on planetary dynamics. An explanation of the interior dynamics of Earth, Venus, Mercury and Mars from the point view of U and Th geochemistry will be given.

## 2. THE GEOCHEMISTRY OF U AND Th

U and Th prefer to combine with $O_2$, $N_2$, $H_2O$ and other volatiles forming relatively low density compounds as shown in Table 1. In addition to the chemical affinities of these low density compounds with silicate and oxides, gravity may also play

**Table 1 The melting point ($^o$C) and density (g/cm$^3$) of U, Th, their compounds and complexes and core related elements(after Bao and Zhang, 1998).***

| Element | Density | Melting point | Compound | Density | Melting point | Compound | Density | Melting point |
|---|---|---|---|---|---|---|---|---|
| Fe | 7.86 | 1535 | $ThO_2$ | 10.0 | 3220 | $UO_2$ | 10.95 | 2800 |
| Ni | 8.90 | 1453 | ThN | 10.6 | 2500 | UCl | 4.86 | 567 |
| U | 19.05 | 1132 | ThS | | 1905 | UN | 14.32 | 2800 |
| Th | 17.70 | 1700 | $ThCl_4$ | | 600 | $UF_6$ | 4.68 | ~0 |
| $K_4ThOX_4 \cdot 4H_2O$ soluble in water | | | $Th(NO_3)_4 \cdot 5H_2O$ soluble in water | | | $UO_3$ | 7.29 | soluble in water |

**\*U and Th can be oxidized very easily in a volatile-rich condition, such as in Earth's surface, the oceanic crust and subduction zone in the shallow mantle. Therefore, there certainly are U, Th compounds and complexes with a much smaller density than these listed in this table. Namely we have not collected the data for all of the U, Th compounds and complexes that can exist naturally.**

an important role in their migration. It is expected that U and Th move up by magmatism and metamorphism processes in a plate tectonic system from the deep mantle to the crust, especially into the granitic continental crusts when they combine with volatiles and water, because of their small density and chemical affinities with silicates and oxides in the crust. Zircon ($Zr_2SiO_4$) is a stable mineral, and can record the concentration variation of U and Th in its parent rocks or magma during its crystallization. There are two kinds of zircons according to their origin: one crystallizes in the magma cooling process and the other during metamorphism, which results in an increase in P and T (Fig. 1). Microprobe data reveal that from the centre to the rim of a magmatic zircon crystal, the U and Th concentration generally increases, but decreases in metamorphic zircons (Bao, 1995). These U, Th zonations were further confirmed by micro-Raman spectra (Bao et al., 1998), and were not formed by U, Th diffusion, which is supported by U-Pb geochronological analyses (Bao, 1995; Bao et al., 1998). Magma cooling is a process of

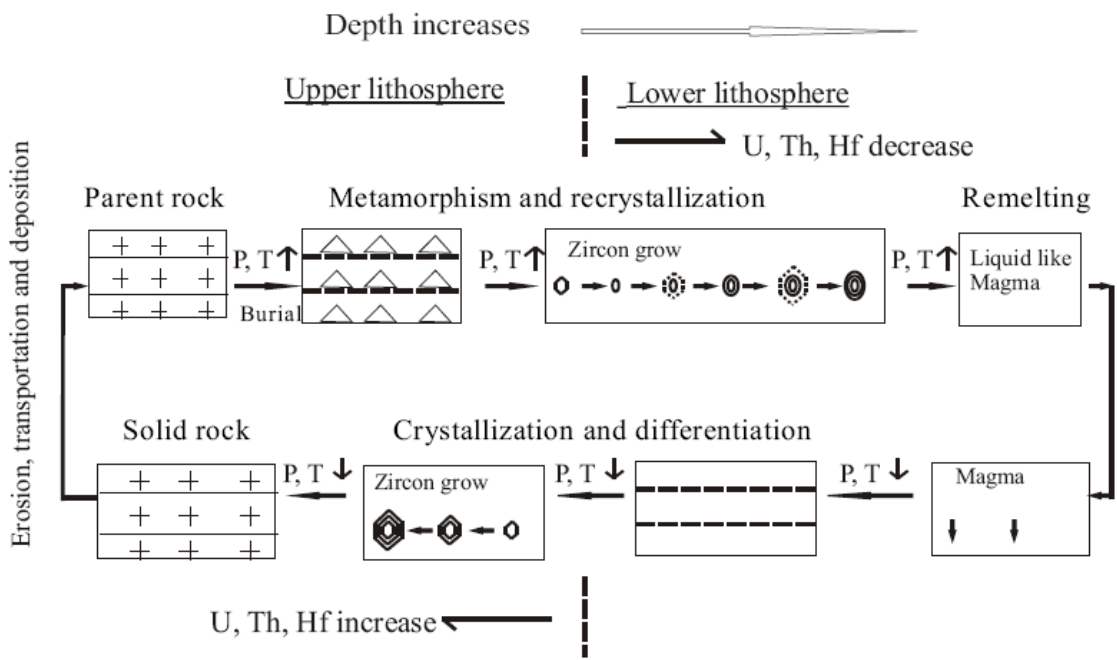

Fig.1 The migration of U, Th and Hf in magmatism and metamorphism. Upper part: metamorphism: Zircon exhibits rounded growth zoning; U and Th decrease from core to rim. Lower part: magmatism: Zircon exhibits euhedral growth zoning. U and Th increase from core to rim or within each small composition zone (after Bao and Zhang, 1998).

temperatures decreasing with time. Therefore, the pattern of U and Th zonation from centre to rim in a zircon crystal indicates that concentrations of U and Th in remnant magma increase with decreasing temperature. Consequently, the concentration of U and Th are the highest at the latest low temperature stage of magma evolution. This is consistent with the observations that U and Th concentration increases from ultramafic (high T), mafic (high to medium T), intermediate (medium T) to felsic rocks (low T). On the other hand, metamorphism is a process of rising temperature and pressure, usually due to an increase in burial depth. Therefore, the reversed pattern of U, Th zonation (relative to magmatic zircon) from centre to rim of a metamorphic zircon crystal, indicates that the concentration of U and Th in their parent rocks decreases with increasing temperature and pressure during metamorphism and they gradually leave from the metamorphic rocks and come back to the shallower crust and surface (low T) by metamorphic heat fluid (Bao and Zhang, 1998; Bao, 1999). Fig. 1 summarizes the migration trends of U and Th during the two processes and shows that U and Th gradually migrate from Earth's high T deep interior to the shallow granitic crust by magmatism and metamorphism processes in a plate tectonic system (Bao and Zhang, 1998). Namely U and Th have been increasing with geological time in the continental crust. Table 2 indicates that the accessory minerals in the young Phanerozoic Granitoids have much higher U and Th than those from the same type of rocks from the old

Table 2 The average contents of U, Th and REE (Rare Earth Elements) in
Precambrian (I) and Phanerozoic (II) granitoids (gram/ton, after Lyafkhovich, 1988)

| Elements | Allanite | Apatite | Zircon | Titanite/sphene | Barringerite |
|---|---|---|---|---|---|
| REE I | 198900.0 | 3514.3 | 3050.0 | 14457.1 | N/A |
| REE II | 210252.8 | 6629.8 | 4895.4 | 17694.1 | N/A |
| U I | 453.0 | 32.5 | 900.0 | 56.6 | N/A |
| U II | 515.2 | 92.8 | 1150.2 | 208.1 | 15 |
| Th I | 7525.0 | 52.8 | 400.0 | 1 | N/A |
| Th II | 7224.4 | 171.8 | 913.3 | 471.2 | 71.5 |

Precambrian (Lyafkhovich, 1988). Granitoids are the main rocks in the continental crust, therefore, the continental rocks become richer in U and Th with geological time. The average concentrations of U and Th are 1.4 ppm U and 5.6 ppm Th in the continental crust (Rudnick and Fountain, 1995), and 0.10 ppm U and 0.22 ppm Th in the oceanic crust (Taylor and McLennan, 1985). This means that the average concentrations of U and Th are 14 and 25 times higher in continental crust than those in the oceanic crust. Furthermore, all continental growth models predict that the volume of continental crust increases through geological time (Abbott et al., 2000 and references therein). Namely the total amount of U and Th in the continental crust increases with geological time, and they must come from the mantle. This supports that U and Th have been migrating from the Earth's deep interior to the continental crust. Therefore, it is suggested that in the upper part of the Earth, U and Th have been migrating up to the continental crust because there are a lot of oxidative volatiles, such as $O_2$, $H_2O$, etc. (Bao and Zhang, 1998; Bao, 1999).

On the other hand, the density of U and Th are much greater than that of Fe, and their melting points are comparable to that of Fe (see Table 1). At the same time, their low valence oxides and compounds also have high melting points and high density values (Table 1). These imply that in an oxygen-poor or reducing environment, U and Th and their low valence compounds will be stable. The high densities present the possibility of their sinking to the deep interior and finally enter Earth's outer core when their parent rock is molten or partially molten (Bao and Zhang, 1998; Bao, 1999).

In addition, our new experimental results indicated that the distribution of U in quenched metal phases is erratic and irregular. This can be seen from the LA-ICP-MS line scans for U content in Fe-10wt% S, Fe-35wt% S and Fe phases shown in Figs. 3-5 in our previous study (Bao and Secco, 2006). This indicates that U may has better solubility in liquid than solid Fe or FeS phases and supports the theory that U can dissolve in liquid planetary cores and Earth's liquid outer core rather than solid metal cores (Bao and Zhang, 1998).

From the discussion above, it can be seen that $O_2$, $H_2O$ and other oxidative volatiles are very important in controlling the migration behavior of U and Th. Experiments show that Si concentration in the metal phase (Fe or FeS) can be used to qualitatively indicate the oxygen fugacity ($fo_2$) of samples in high P, T experiments (Kilburn and Wood, 1997). The new experimental results presented in our previous studies on pure Fe samples indicate that with increasing P, Si in the metal phase increases, which means that $fo_2$ decreases with increasing P (Bao et al., 2006; Bao and Secco, 2006). In addition, increasing T also decreases $fo_2$ (Righter and Drake, 2006). This well explains why an increase in P and T causes an increase in U solubility in Fe and FeS phases as presented in previous studies (Bao et al., 2006; Bao and Secco, 2006). A positive correlation between the $fo_2$ and $H_2O$ content of mantle rock is also confirmed by high P, T experiments (McCammon et al., 2004). These further support the earlier findings that in reducing and $H_2O$-poor conditions, U and Th are prone to enter Earth's liquid outer core (Bao and Zhang, 1998; Bao, 1999).

In the mantle, increase in depth is accompanied by an increase in P and T, which in turn means that $fo_2$ also decreases according to the correlation between $fo_2$ and P, T as discussed above. Therefore the lower mantle and core are at a more reducing condition than the crust and upper mantle (McCammon et al., 2004). In other words, there are few oxidative volatile elements surviving within the lower mantle and core. Therefore, most of the U and Th in the mantle and core must have existed in a state of metal or low valence compounds. In this case, U and Th will stay in the lower mantle and gradually sink to the base of the mantle, and part of them may have been sequestered into the liquid outer core by core-mantle chemical interaction. Since the concentration of oxidative volatile compositions are different among different planets as discussed in the following, therefore, the distribution of U and Th in different planets are also expected to be different, which will lead them to have different geodynamic and evolution histories.

## 3. IMPLICATION FOR TERRESTRIAL PLANETS
3.1. The distribution of oxidative volatiles in terrestrial planets

The distribution of volatiles (including oxidative volatiles) among planets is mainly controlled by two factors: their heliocentric distance (Lewis, 1974a, 1984) and size (Ahrens, 1993). According to the planetary accretion hypothesis (e.g. Fegley, 1999), planets have formed from a protoplanetary disk surrounding the proto-Sun. In the protoplanetary disk, the mid-plane is approximately 400 $^oC$ higher than the surface within the range from Mercury, 0.397 astronomical unit (AU), to Mars, 1.524 AU (Boss, 1998), and the P and T gradients in the mid-plane were controlled by the heliocentric distance as shown in Fig. 2. The closer towards the proto-sun, the higher the temperature becomes in the mid-plane. Within 1.0 AU, this T in the protoplanetary disk was greater than 323 $^oC$ (600 K). Therefore the volatile elements in the inner part of the terrestrial planetary belt (< 2.7 AU) could have been mostly vaporized and pushed to the outer reaches of the disk by solar wind (Beaty et al., 2005). This is consistent with the astronomical observations that volatile concentration increases with heliocentric distance. Mercury, the closest planet to the sun, has the largest proportion of a metallic core; its metallic core occupies 64% of Mercury's weight and 42% of its total volume (Strom and Sprague, 2003). Mid-infrared spectroscopic measurements shows that Mercury's silicate mantle (this is its outermost layer since it does not have an Earth-like crust) has the lowest amount of FeO compared to other terrestrial planets (~3%, e.g. Emery et al., 1998, or < 0.1%, Burbine et al., 2002), suggesting that almost all of its Fe has differentiated into its core (Strom and

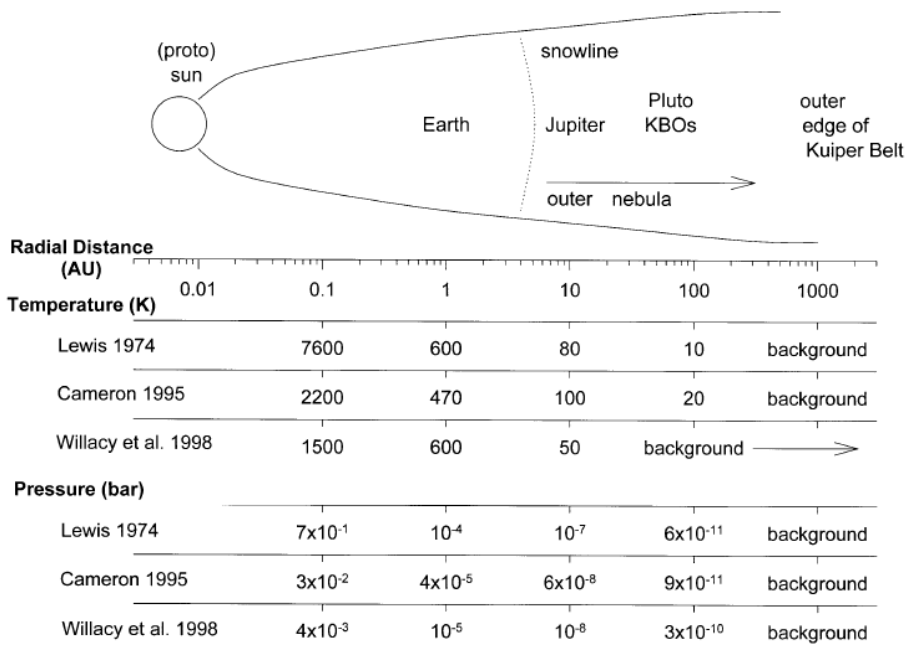

Fig. 2 A schematic diagram comparing the radial variation of temperature and pressure for three solar nebula models (after Fegley, 1999).

Sprague, 2003; Righter and Drake, 2006). In other words, this planet has the lowest oxygen content. Similarly, it was found that the S content in the Venusian atmosphere is also very low in comparison with Earth. Therefore, Lewis (1974b) argued that there should not have any S in the original building materials of Venus. This implies that the temperature in the protoplanetary disk in the feeding zone of Venus was greater than the vaporization temperature of S or FeS at the relevant pressures in the proto-planetary disk. Recent research indicates that planets usually form from a narrow feeding zone in the proto-planetary disk (Righter and Drake, 2006). Therefore, it is expected that the original building materials of Mercury and Venus should be largely devoid of S, and by extension possibly other volatiles as well. The present atmosphere of Venus may have formed from later cometary impacts (Lewis, 1974b). Therefore, it can be concluded that the volatile content goes as: Mercury < Venus < Earth < Mars < asteroid belt < outer planetary belt from the view of their original building materials.

However, the volatile content of a planet is also controlled by another factor - its size (Ahrens, 1993). Gravitational accretion by impact is the main mechanism of planetary formation. Therefore, the atmosphere and water on a planet were formed by impact-induced dehydration of water-bearing minerals in planetesimals (e.g. Ahrens, 1993). However, if a planet is too small, this kind of impact can also drive off these volatile compositions so that they are no longer held by the gravity field of the proto-planet. Calculation results indicate that the impact energy needed for Earth to lose its entire atmosphere is ~$10^{38}$ ergs; for Venus $10^{36}$ - $10^{37}$ ergs and for Mars $10^{34}$ - $10^{36.5}$ ergs (Ahrens, 1993). This means that even a relatively small impact can drive off the Martian atmosphere, including the volatile compositions in its interior. Therefore, although Martian original building materials may be richer in volatile compositions than those of Earth, it may have lost them during its accretion and formation stages, or at a later stage (see discussion later). Mars may have been a dry planet in its interior throughout a long history. The limited surface water may have come from the addition of later cometary impacts, as suggested in the case of Venus (Lewis, 1974b). Mercury would have lost its very limited volatile compositions because of its much smaller volume and mass compared with Mars.

Hydrodynamic escape can also play an important role in losing volatiles when there is a later core formation in a planet, where the incoming volatiles mixing with free Fe allows water to oxidize Fe, which would release hydrogen into the atmosphere and produce a hydrodynamic escape (Beaty et al., 2005). However, the Earth's core formed within a very short time span of 30 Ma after Earth's accretion (Jacobsen, 2005). Hafnium-tungsten isotopes in meteorites also suggest that the Martian core formed within 7-15 Ma after the formation of Mars (Kleine et al., 2004). Hence, hydrodynamic escape may not be an important mechanism in the release of volatiles for Mars. For Mercury and Venus, there is insufficient data to estimate the time of their core formation. However, according to the negative correlation between FeO contents in the mantle and heliocentric distance of the terrestrial planets (Righter and Drake, 2006), the mantles of Mercury and Venus should have less FeO than those of Earth and Mars, suggesting that hydrodynamic escape may not have played an important role in releasing their volatiles. Thus the heliocentric distance and size of a terrestrial planet are the most important factors in controlling its volatiles, including these oxidative volatiles.

In summary, Earth's interior is rich in volatile components (~0.01%, Beaty et al., 2005), and these include a large amount of oxidative volatiles and water. Mercury certainly is a dry planet. Venus should contain less oxidative volatile components than Earth because of its smaller heliocentric distance, and smaller mass and volume. Mars may be dry as a result of one or more large impacts.

3.2. An explanation of the differences in geophysical features between Earth and Venus

Venus is similar in heliocentric distance, size, density and composition with Earth (Basaltic Volcanism Study Project, 1981). Moreover, it is also expected to have concentrations of the heat-producing elements (U, Th, K) in its interior similar to the concentrations within the Earth. Measured surface concentrations of these elements on Venus (Surkov et al., 1987) have confirmed this basic hypothesis. However, it has different geophysical features and a volcanic and tectonic history differing in important respects from that of Earth. These include: 1) Recent estimates of the tidal Love number from Magellan's orbit by Konopliv and Yoder (1996) suggest that the Venusian core may be liquid, but it appears to lack an appreciable Earth-like dipolar magnetic field. This difference is a long-standing puzzle that cannot be explained by the planet's slow rotation (Russell, 1980); 2) Unlike Earth, Venus does not have a plate tectonic system. Instead, it has a lot of volcanic plains. The ages of most of these volcanic plains (which occupy more than 80% of the total surface area) are similar and their formation is thought to have occurred in a short time interval of less than 100 Ma (Nimmo et al., 1998; Turcotte et al., 1999; Basilevsky, 2002). Therefore, global catastrophic resurfacing is expected to have occurred. Some researchers considered that this kind of resurfacing event happens periodically on Venus (Turcotte et al., 1999). Regardless of the uncertainty of their causes and nature (Nimmo et al., 1998), it is reasonable to question the underlying cause(s) of these basic differences between the Earth and Venus. Nimmo et al. (1998) considered the main reason is that Earth possesses water and Venus does not. However, the fundamental importance of water to processes on Earth has not been fully recognized (Nimmo et al., 1998).

Based on previous studies (Bao and Zhang, 1998; Bao, 1999) and new experimental results presented in the new studies (Bao et al., 2006; Bao and Secco, 2006), it is proposed that the different migration behaviors of U and Th within the two planets may be due to their differences in water and oxidative volatile content. Each terrestrial planet is expected to have a significant amount of U and Th in its metal Fe core. However, Earth and Venus have different contents of water and oxidative volatiles within their interiors (e.g. Lewis, 1974a, 1974b). This would have led them to have different U and Th distributions, and in turn resulted in different planetary dynamics (Bao and Zhang, 1998; Bao, 1999). In Earth's early stage, the core had a large amount of U and Th. However, the relatively rich oxidative volatile components in its building materials could have been partially sequestered into its metal core during the core formation. Therefore, when the inner core begins to crystallize, these volatiles, such as O (Rubie, et al., 2004), H (Wood, 1997), S and C (e.g. Ishida et al., 1999) would be released and enter the liquid outer core (e.g. Buffett et al., 2000). They would first oxidize the U and Th in the liquid outer core and form smaller density U, Th compounds and complexes since U and Th are electropositive (Urey, 1955) and therefore volatiles combine more easily with U and Th than with Fe in the outer core. They are then expected to migrate up to the mantle and crust by hot super-plumes as indicated in Fig. 3. High P and T experimental results (Murukami et al., 2002) have shown that the representative minerals in the lower mantle can contain 5 times the amount of water at the Earth's surface, and therefore volatile ions and water may have been transported into the lower mantle by the hydrated subducted slabs (Murukami et al., 2002) or cold super-plumes (Ishida et al., 1999). Consequently there is a possibility that the cold super-plume with these volatile compositions or oxygen-rich materials (Walker and Walker, 2005) can be moved down to the core in compensating lost material of the core at the hot-plumes (Bao and Zhang, 1998; Bao, 1999; Ishida et al., 1999; Walker and Walker, 2005). These water and oxidative volatile compositions will help to oxidize more U and Th in the outer core, and then convert them to U, Th compounds and complexes with a lower density (Bao and Zhang, 1998; Bao, 1999, see Table 1). At the same time, U and Th can be important contributors to the formation of an asthenosphere, or U, Th rich sphere (Bao and Zhang, 1998; Bao, 1999). Since heat produced by U and Th can lower the viscosity there, consequently the overlying oceanic plates can shift and migrate gradually to the subduction zone and enter the mantle again. Therefore, the outer core, hot super plume(s), asthenosphere or U, Th rich sphere and subduction zone (or cold super plumes) have formed a circulation system inside the Earth as shown in Fig. 3. The heat from U and Th in the core may be the main energy sources to drive this circulation system to work, and the abundant water and oxidative volatiles inside the Earth makes it workable (Bao and Zhang, 1998; Bao, 1999). Within this system, U and Th are gradually moved into the asthenosphere, and then into the continental crust by combining with more water and volatiles coming down from the oceanic crust at subduction zones. Furthermore, it may be the loss of U and Th in the core in this way that the inner core could begin to crystallize at Earth's early stage. The other direct results are the formation of granitic continental crust with a high content of U and Th and the plate tectonic system (Bao and Zhang, 1998; Bao, 1999).

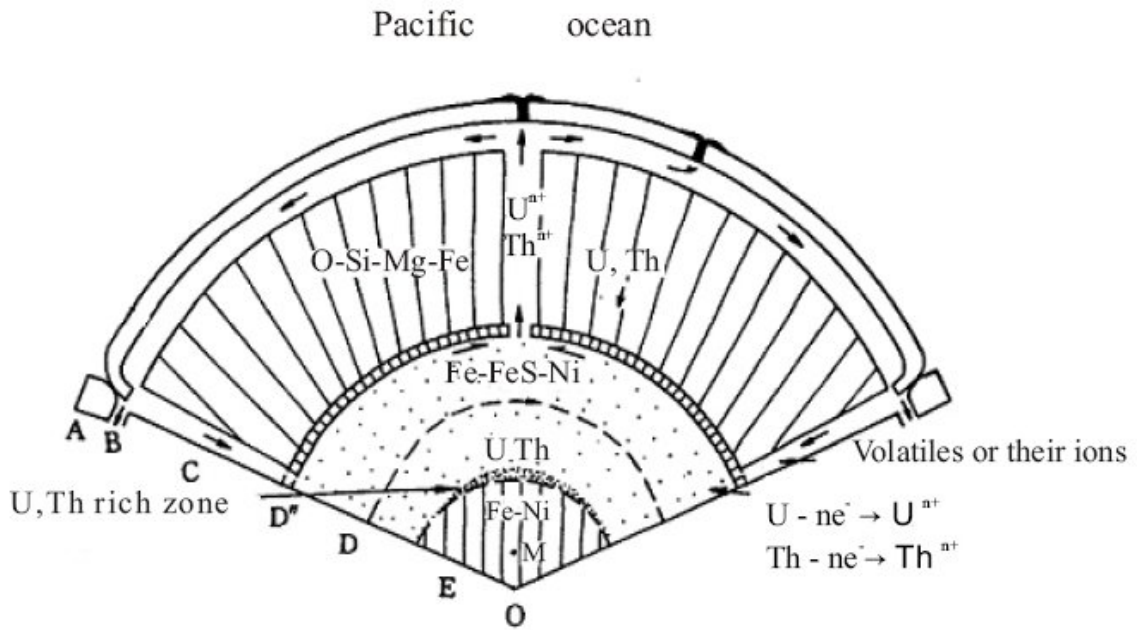

**Fig. 3 The distribution of U and Th in the outer core and its influence on the formation of deep mantle plumes and subducted lithosphere plates in the Pacific Ocean. A-lithosphere; B- U,Th richer sphere or asthenosphere; C- mantle; D''- core-mantle boundary; D- outer core; E- inner core; O- center of Earth; M- center of geomagnetic field and thermal convection; the black points represent the relative concentrations of U and Th (after Bao, 1999).**

The assumption that the asthenosphere is rich in U and Th (or there is a U, Th rich sphere in the asthenosphere position), proposed by Bao and Zhang (1998), has been used to explain the U source in several uranium ores and rock bodies (e.g. Wu and Chen, 1999; Jiang et al., 2000; 2002a; 2002b; 2004; Wu et al., 2005). This implies that there is a need for some amount of U and Th in the asthenosphere. Kimberlites are usually rich in U, Th and other incompatible elements (Claude-Jean and Michard, 1974), and the source region of kimberlites is considered to be located within the asthenosphere (Le Roex, 1986), which directly supports the assumption that the asthenosphere is rich in U and Th. Therefore, the heat released from U and Th in the asthenosphere may be the key to cause the rocks there to melt partially, which is the pre-requisite to make the overlying oceanic lithosphere plate movable and consequent plate tectonics workable.

Other evidence include: a) the Os isotope study indicates that super plume can bring the core materials from the outer core to the crust (Brandon et al., 2003); b) Seismic tomographic technique has imaged a hotter than average structure that originate from the outer core, through super plumes and is reflected by the lithosphere and horizontally going through the asthenosphere (Garnero et al., 1998, Romanowicz and Gung, 2002); c) Seismic tomography also confirmed that subducted oceanic lithosphere (seismic "cold" region) can penetrate into the lower mantle (e.g. Tanimoto and Lay, 2000; Courtillot, et al., 2003). Connecting these facts together, we can conclude that an internal circulation system may possibly have existed within Earth. A hotter than average structure from core-mantle boundary, super plumes to the asthenosphere may partly be produced by the heat released by U and Th in the circulation system since U and Th concentrations of basalts from the deep plumes (OIB) are about 40 times higher than these from MORB (Winter, 2001; Sun and McDonough, 1989).

Venus has a lower Na content than Earth (Palme, 2002) and this indicates that it has lower volatile components, which is consistent with the conclusion reached in section 3.1. Consequently, the U and Th within Venus will have less possibility to combine with oxidative volatiles and water, and will exist as metals or simple compounds with a high density (see Table 1). This will lead them to gradually sink to the core of Venus from the mantle and prevent the formation of an Earth-like asthenosphere. Therefore, on the basis of this explanation, it is unlikely that a plate tectonics system has developed on Venus.

When U, Th reach a large concentration in the core, they will produce great amounts of heat, and this heat will trigger a catastrophic resurfacing event (Nimmo et al., 1998, Basilevsky, 2002, Turcotte et al., 1999) by strong super-plume activities as shown in Fig. 4. In this process, the U and Th in the core may be partly moved out from the core to the mantle, even to the surface. Venus will then enter a relatively quiescent period like the present Venus after a large amount of heat was released during the resurfacing events around 700 Ma (Nimmo et al., 1998, Basilevsky, 2002, Turcotte et al., 1999). However, with the increase of mantle temperature, U and Th in the solid mantle may sink again into the core, and they become the energy source to lead to the next catastrophic resurfacing event. More U and Th in the core of Venus will prevent the formation of an Earth-like

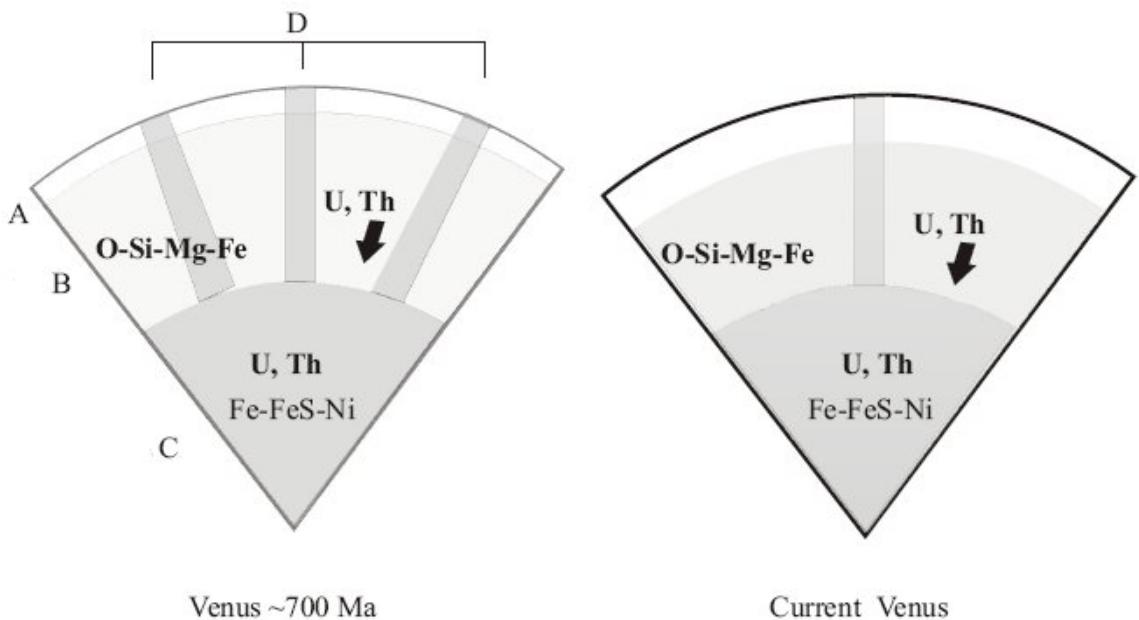

**Fig. 4 Schematic representation of the global catastrophic resurfacing event around 700 Ma (left) and present relatively quiescent period (right) of Venus.   A- Crust; B- Mantle; C- Core; D- Plume**

inner core, which also prevents the formation of an Earth-like dipole magnetic field.

3.3. An explanation of the dipole magnetic field of the Mercury

Mercury is the closest planet to the Sun; therefore, its original composition most likely came from the innermost part of the solar nebula (Strom and Sprague, 2003; Righter and Drake, 2006). Consequently, it may have the most reducing materials and the smallest amount of volatiles (Burbine et al., 2002). It is the smallest planet in the solar system (its volume is only about 6% that of the Earth), but it has the largest ratio of Fe core with a total diameter of 0.75. The core of Mercury constitutes about 42% of its volume. In contrast, Earth's iron core is 54% of the total diameter but constitutes only 16% of the volume (Strom and Sprague, 2003). Mercury's large Fe core shows that post–accretion vaporization or giant impact stripping removed a large fraction of the silicate mantle from a once larger proto-Mercury (Strom and Sprague, 2003). A thin silicate mantle of 639 km thick will make it lose its internal heat easily. It is expected that its metal core solidified within a very short time after its accretion. However, it has an Earth-like dipole magnetic field, although its intensity is only 1/100 of Earth's (Strom and Sprague, 2003).

So far, there are two hypotheses to explain the origin of Mercury's dipole magnetic field. The first one considered that there is a liquid outer core because of the presence of ~ 0.2 – 5 wt% sulfur, which lowers the melting point of the core material (e.g. Strom and Sprague, 2003). However, if Mercury formed in its current position (Righter and Drake, 2006), then it should lack S (Strom and Sprague, 2003). The building materials of Venus may have no S (Lewis, 1974b); hence, it is difficult to imagine that the building materials of Mercury has significant S with a much smaller heliocentric distance than Venus according to the discussion above.

The second hypothesis (e.g. Strom and Sprague, 2003) assumed that the dipole magnetic field is a remnant one, and its core is now completely solidified with an S content of less than 0.2% (e.g. Strom and Sprague, 2003). This small amount of S may be consistent with the origin of Mercury, but a remnant field requires a thicker than 30 km magnetic mineral layer, which may be impossible from a geochemical or petrologic standpoint, and also conflicts with geological observation, such as low FeO in its mantle (Strom and Sprague, 2003).

According to the experimental results of pure Fe presented in previous study, U can dissolve in pure Fe phases and its solubility increases with P and T (Bao et al., 2006). Therefore, as with Venus, Mercury may also have had a lot of U, possibly also Th (due to their similarities), in its core but they cannot be moved up to the surface due to an absence of water and oxidative volatile elements. Because of Mercury's small volume and thin rock mantle, the release of its internal heat energy should be very fast compared to Earth and other planets. Therefore, the crystallization of its inner core would have necessarily begun at an early stage. However a significant amount of U and Th in the core would prevent its complete solidification, and leave a remnant liquid outer core inside Mercury. The convection of the liquid outer core is responsible for the creation of a dipole magnetic field (Stanley et al., 2005). On the other hand, the existence of a dipole magnetic field on Mercury can be a piece of evidence to support the conclusion that U, and possibly Th, can enter terrestrial planetary cores in significant amount. Like S, K is a volatile element, and it is unlikely for it exist in Mercury's original building material at such a small heliocentric distance. Accordingly, U (and Th) in planetary cores may have also played an important role in Mercury's evolution.

3.4. A view from U geochemistry on the interior dynamics of Mars

Mars has very large volcanic rises. The Tharsis, with a range of ~5000 km of topographic uplift consisting of several high volcanoes (Olympus Mons, ~25 km high), is the tallest mountain in the solar system. These volcanoes may have originated 3.7 Ga ago, but may still be active today (e.g. Nimmo and Tanaka, 2005). This indicates that Mars has had great internal energy, which may still be present today. The next most obvious feature on the Martian surface is the hemispheric dichotomy structure. The crust in the southern hemisphere consists of older (>3.7Ga, formed during Noachian period) and less dense highlands with a composition of andesite or basaltic andesite (e.g. Nimmo and Tanaka, 2005). They may be rich in rare-earth and radiogenic elements (e.g. Nimmo and Tanaka, 2005). On the contrary, the crust in the northern hemisphere consists of younger (<3.7Ga) lowlands with basaltic composition (Nimmo and Tanaka, 2005). The southern lowland was resurfaced by later Hesperian (3.7-3.5 Ga) and Amazonian (3.5-today) events (Nimmo and Tanaka, 2005). The formation age of this dichotomy structure was considered to be around 4.5 Ga (e.g. Solomon et al., 2005) or 4.12 Ga (e.g. Nimmo & Tanaka, 2005). However, it is only a relative age based on the impact crater size-frequency distributions, and there is great uncertainty (Nimmo and Tanaka, 2005).

Mars possesses no internal magnetic field and most of its crustal rocks have not been magnetized. However, in limited regions in the south, the amplitude of the strongest magnetization is comparable to or greater than the highest common values in Earth's crust, which implies that a dynamo inside Mars was present in its early stage but must have been absent after 4 Ga (e.g. Fei and Bertka, 2005; Solomon et al., 2005, Nimmo and Tanaka, 2005).

A dry Martian mantle since its core formed was pointed out by Dreibus and Wänke (1985, 1987) and Carr and Wänke (1992) based on the chemical composition of Martian meteorites. This is followed by Reese and Solomatov (2002), who emphasized a relatively dry Martian mantle due to the fact that the water content of SNC meteorites is an order of magnitude lower than the typical MORB on Earth. A reduced and Fe-depleted Martian mantle source region was also found by a Martian meteorite study (Ghosal et al., 1998). Herd et al. (2002) estimated that the log $fo_2$ of the Martian mantle is in the range of -3 to -1 at the quartz-fayalite-magnetite (QFM) buffer (being equal to -1 to 0 for iron-wustite (IW) buffer). Compared with the $fo_2$ of Earth's mantle: ~IW +3 (Stolper and McSween, 1979), the Martian mantle is rather reduced, and may have had limited amounts of water in its early stage (Jones, 2004). As summarized by Halliday et al. (2001), Mars is expected to be more depleted in volatiles than Earth, although there are doubts from other geochemical evidences (McSween et al., 2001).

The latest evidences seem to strengthen the view of a dry Martian interior. The OMEGA (Infrared mineralogical mapping spectrometer in France) reflectance spectra indicate that olivine and pyroxene dominate the Martian surface. Only limited areas, such as the old unburied cratered units, have hydrated phyllosilicates and sulfates, but no carbonates (Mustard et al., 2005; Bibring et al., 2005). Moreover, in the volcanic outflows from Nili Patera, no hydrated minerals were detected (Bibring et al., 2005). Beaty et al. (2005) concluded that the content of Martian interior volatiles is 2000 times less than that of Earth. These further indicate that Mars has a dry interior, which is consistent with the conclusion of section 3.1.

According to Fei and Bertka (2005), the Martian metal core may contain 14.5% S. This is much higher than that of Earth, and can lead to more U entering the Martian core according to the experimental results of our previous study (Bao et al., 2006; Bao and Secco, 2006). Therefore, there is greater possibility to have more U (> 10 ppb) (and possibly Th) in its core (Bao et al., 2006; Bao and Secco, 2006). At the same time, more oxygen and oxidative volatile elements may be dissolved into the metal core during its formation because of the high content of oxygen and other volatiles in the Martian building materials. These light elements could have combined with U and Th and led them to migrate to the crust by the Earth-like planetary interior circulation system described in section 3.2. This would lead to the cooling down of the core and crystallization of a solid inner core. This also would start a Martian dynamo and a similar plate tectonic system in its early stage, and create a dipolar magnetic field. This is supported by the latest results of Connerney et al. (2005), who found several additional evidences, including lineal magnetic imprints and similar transform faults that support plate tectonic development on the young Mars. Like the situation on Earth, this internal circulation system also could create an andesitic or basaltic andesitic crust (only in the southern hemisphere, the original crust in northern hemisphere may be destroyed by impact event(s) as discussed next) with lower melting points and a greater concentration of heat-producing elements.

However, the special hemispheric dichotomy structure and the wide–ranging demagnetization on Martian surface indicate that there may have been one (Wilhelms and Squyres, 1984) or several (Frey and Schultz, 1988) large impact events in its early stage. After these giant impact events, the special hemispheric dichotomous structure was formed and the impacts may have also driven off most or all of the oxidative volatile compositions inside the planet, which would result in cessation of the interior circulation system (Bao and Zhang, 1998, Bao, 1999). The Martian dynamo would have ceased with the end of this internal circulation. Afterwards, the heat released by U and Th in the core can be only moved up by super plumes, such as the volcano system of the Tharsis rise.

This is similar to the present Venus described in section 3.2. Recently Yoder et al. (2003) confirmed by the detection of solar tide that Martian metal core is at least partially liquid. This supports that Mars still has plenty of U and possibly Th in its core.

**Acknowledgements**

I thank Dr. R. A. Secco for his support, discussion and constructive criticism; Dr. R.V. Murthy, Dr. S. Shieh, and Dr. R.A. Flemming for their comments and suggestions.
.